\documentclass[11pt]{article}

\usepackage{geometry}                
\usepackage[parfill]{parskip}    
\usepackage{graphicx}
\usepackage{amssymb}
\usepackage{amsmath}
\usepackage{epstopdf}
\usepackage{bm}
 \usepackage{setspace}

\newtheorem{theorem}{Theorem}
\newtheorem{lemma}[theorem]{Lemma}
\newtheorem{example}[theorem]{Example}

\newenvironment{proof}{\noindent {\em Proof}\\}{\hfill $\Box$\\}

\newcommand{\bH}{\mathbf{H}}
\newcommand{\bI}{\mathbf{I}}
\newcommand{\bP}{\mathbf{P}}
\newcommand{\bQ}{\mathbf{Q}}
\newcommand{\bR}{\mathbf{R}}
\newcommand{\bX}{\mathbf{X}}
\newcommand{\bY}{\mathbf{Y}}

\newcommand{\bp}{\mathbf{p}}
\newcommand{\bq}{\mathbf{q}}
\newcommand{\bu}{\mathbf{u}}
\newcommand{\bv}{\mathbf{v}}

\newcommand{\be}{\begin{equation}}
\newcommand{\ee}{\end{equation}}

\newcommand{\bLambda}{\mathbf{\Lambda}}
\newcommand{\bt}{\mathbf{\tau}}
\renewcommand{\th}{\mathrm{th}}
\newcommand{\shalf}{{\mbox{\tiny$\frac{1}{2}$}}}

\newenvironment{rmatrix}{\left(\begin{matrix}}{\end{matrix}\right)}

\title{Hadamard Phylogenetic Methods and the $n$-taxon Process}

\author{David Bryant\thanks{Dept. Mathematics, University of Auckland. email: {\tt bryant@math.auckland.ac.nz} web: {\tt http://www.math.auckland.ac.nz/\~{}bryant}}}

\begin{document}
\maketitle

\begin{abstract}
The Hadamard transform of \cite{Hendy89a, Hendy89} provides a way to work with stochastic models for sequence evolution without having to deal with the complications of tree space and the graphical structure of trees. Here we demonstrate that the transform can be expressed in terms of the familiar $\bP[\bt] = e^{\bQ[\bt]}$ formula for Markov chains. The key idea is to study the evolution of vectors of states, one vector entry for each taxa; we call this the $n$-taxon process. We derive transition probabilities for the process. Significantly, the findings show that tree-based models are indeed in the family of (multi-variate) exponential distributions.
 \end{abstract}


\section{Introduction}

\subsection{Stochastic models and the Hadamard transform}

Stochastic models for sequence evolution now play a part in most phylogenetic analyses. Given a tree, and a set of branch lengths, the models determine a probability distribution for the patterns of nucleotides/amino acids observed at a site in the alignment (the {\em site patterns}). The real difficulty of these tree-based models is that they are, indeed, based on a tree. The graphical structure of the tree is intrinsic to the probability formulae. Here, as in many other contexts, spaces of graphical structures are difficult to work with, both for the statistician and computer scientist. Indeed the {\em state-of-the-art} optimisation methods go little beyond local search techniques.

The Hadamard transform \cite{Hendy89} circumvents many of these issues. Each tree is encoded as a vector called a {\em split vector} or {\em spectrum} $\bq$. Hendy and Penny showed that there is a general formula taking a spectrum to the vector of site pattern probabilities for the tree:
\be \bp = \bH^{-1} \exp(\bH \bq).\ee 
The matrix $\bH$ is a Hadamard matrix, see Section~\ref{sec:singlebranch}. This formula works for all trees and, once the tree has been coded as a vector, the tree structure plays no further part in the computation. Thus the Hadamard transform provides a model into which all tree-based models naturally nest. Refer to \cite{Swofford96, Felsenstein04} for excellent introductions to Hadamard transform methods.

The transform is a useful tool for many theoretical investigations. However there are also important practical advantages of the approach. For example:
\begin{enumerate}
\item It provides a way of searching (or potentially, sampling) tree space that does not involve passing from individual tree to individual tree. One can invert the Hadamard transform to construct a spectrum $\bq$ from the observed pattern probabilities and then use $\bq$ to infer a tree.
\item Because the transform provides a model that generalises trees it can be used to test the hypothesis ``does this data actually come from a tree?'' An analysis of this sort does not need to be hypothesis driven: phylogenetic network software like SplitsTree \cite{Huson06} and Spectronet \cite{Huber02} allow one to visualise a spectrum $\bq$ and see where it violates tree based models.
\end{enumerate}
 
There have been many reformulations of the original Hadamard transform formula. Early proofs of the transform were based on an interpretation in terms of path sets \cite{Hendy89a, Hendy89}; these were recently extended in \cite{Hendy05}. The transforms were recast in terms of Fourier transforms on Abelian groups \cite{Evans93,Steel92,Szekely93,Szekely93a}. Bryant \cite{Bryant05a} used this algebraic machinery to show that the transform can be understood in terms of evolutionary models on phylogenetic networks. Sturmfels and Sullivant \cite{Sturmfels05} view the transform as a {\em change of coordinates} and, like \cite{Evans93}, use it to study invariants on the phylogenetic tree models.\\

 There are two important limitations of the Hadamard transform. The first is the running time: a full Hadamard transform takes exponential time: current analyses are limited to a maximum of $30$ taxa. This could be remedied using approximations (such as distance based methods like NeighborNet \cite{Bryant04}) or Monte-Carlo strategies.
 
 The second limitation is the restriction on the substitution models with the Hadamard transform to {\em group-based} models. For nucleotide data, this means that one can only use subsets of the K3ST model (see Section~\ref{sec:treemodels}). This restriction is probably the most important barrier to use of the transform. It was one of the motivations behind the reformulation of the transform outlined in this paper. The problem of how to remove this restriction is still open.
 
 \subsection{Contribution of this paper}

In this paper we derive a new formulation of the Hadamard transform. Our proof that the transform works does not use path sets or Fourier transforms, at least not explicitly. The transform is established using basic matrix analysis that does not go much beyond the tools the are routinely used in phylogenetic analysis.

The key idea in the current paper is the {\em $n$-taxon process}. Consider a phylogenetic tree with times marked. At any particular time, every taxa has a unique ancestral lineage and this lineage has a unique state. Let $\bv_t$ denote the vector of ancestral states, so that $\bv_t[i]$ is the state of the ancestor of taxa $i$ at time $t$. The $n$-taxon process is the continuous time Markov chain that describes the evolution of these vectors over time.

We derive the transition probability matrix $\bP[\bt]$ for this process for a given vector $\bt$ of branch lengths. It is simply the exponential 
$\exp(\bQ[\tau])$ of a linear combination of rate matrices for the branches. The formula for the Hadamard conjugation falls straight out of the formula for $\bP[\bt]$, recast in appropriate notation. In fact the vector $\bq$ as defined in \cite{Hendy89} is simply the first column of the rate matrix $\bQ[\tau]$.

The entry $\bP[\bt]_{\bu \bv}$ gives the probability that the final state is $\bv$ (these are the observed states for each taxa) given that the initial vector of states is $\bu$. In the standard model for phylogenetic analysis that states for the initial vector would be all the same, corresponding to the fact that in a phylogeny the root is ancestral to all taxa. The same may not apply to problems from population genetics.

In this paper we examine only two models, the binary symmetric model and the K3ST model. In practise, these are the only two models used with the Hadamard conjugation. The results here could be generalised to general group based models \cite{Evans93,Szekely93, Bryant05a}. 

\section{From tree based models to the $n$-taxon process}

\subsection{Tree based models} \label{sec:treemodels}

We begin with a brief outline of the standard models used for sequence evolution; for more details  and extensive references, see \cite{Bryant05,Felsenstein04,Swofford96}.

As usual,  we make the assumption that different sites in a sequence  evolve independently from each other, so we can consider just the evolution of a single site. A state is drawn at the root from a fixed distribution $\pi$. The evolution of the site then proceeds along the branches from the root to the leaves of the tree. Along each branch, substitutions occur according to a continuous time Markov chain, the chain specified by its {\em (instantaneous) rate matrix} $Q$.

\begin{figure}[htb] 
   \centering
   \includegraphics[width=5in]{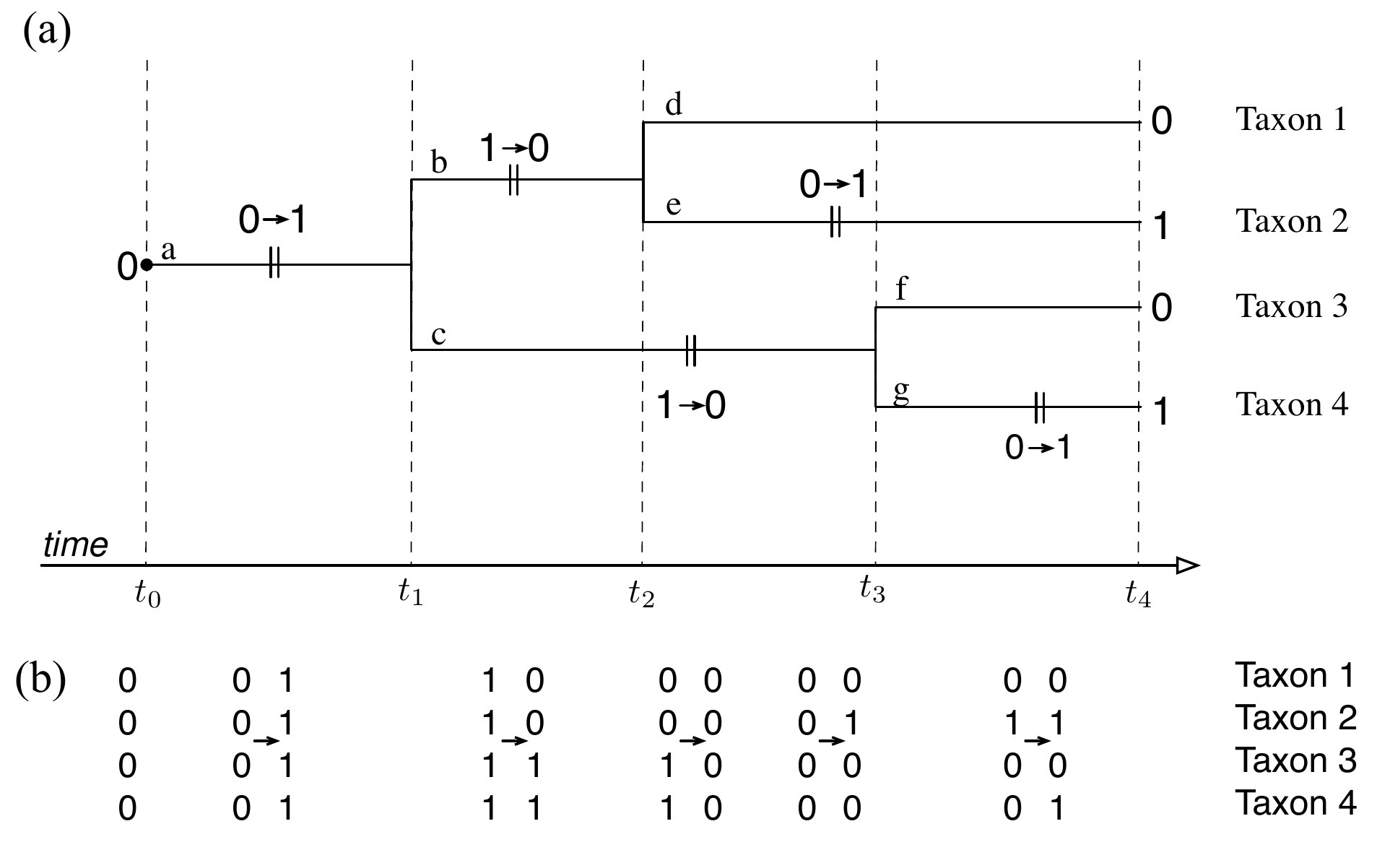} 
   \caption{\small (a) An example of state evolution on a tree under the binary symmetric model. The horizontal axis is proportional to time. The state $0$ was drawn at the root, and substitutions occurred on edges a,b,c,e,g, giving the pattern 0101 at the leaves. (b) The corresponding $n$-tuple process. At each stage, the value of the process gives the ancestral states for each taxon. This changes five times from left to right.}
   \label{fig:modelseq}
\end{figure}

\begin{figure}[htb] 
   \centering
   \includegraphics[width=1in]{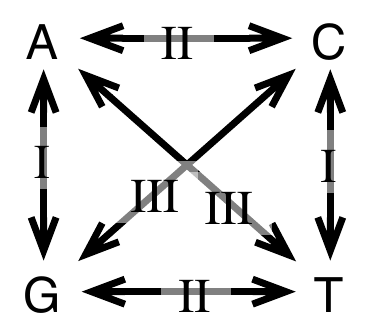} 
   \caption{The three transition types under the K3ST model}
   \label{fig:K3ST}
\end{figure}

We consider only two choices for $Q$, corresponding to the {\em binary symmetric model} and the {\em K3ST} model. They have respective rate matrices
\be  Q_{(2)} = \begin{rmatrix} -1 & 1 \\ 1 & -1 \end{rmatrix}, \hspace{1cm} 
Q_{(4)} =  \begin{rmatrix} 
-\alpha-\beta-\gamma & \alpha & \beta & \gamma \\
 \alpha & -\alpha-\beta-\gamma & \gamma & \beta \\
\beta & \gamma & -\alpha-\beta-\gamma & \alpha \\
\gamma & \beta & \alpha & -\alpha-\beta-\gamma
\end{rmatrix}
.\ee 
The values $\alpha,\beta,\gamma$ are positive reals chosen so that $\alpha+\beta+\gamma=1$. These parameters control the rates of three types of substitution as represented in figure~\ref{fig:K3ST}. Type I substitutions correspond to DNA transitions; types II and III are types of transversion. Different sub-models are obtained by making different rates equal to each other.

Both the binary symmetric model and the K3ST model have uniform stationary distributions. This is used for the distribution $\pi$ at the root.

\begin{example}
We illustrate the binary symmetric model on a four taxa tree in figure~\ref{fig:modelseq} (a). The root distribution is uniform: in this case $0$ was drawn (with probability $1/2$). Substitutions occurred on edges a,b,c,e,g, giving the pattern 0101 at the leaves. 
\end{example}

Let $P_{ij}(t)$ denote the probability that  the state at the end of a branch of length $t$ is $j$ given that the state at the beginning of the branch is $i$. These {\em transition probabilities} are given by the matrix exponential
\be P(t) = \exp(Qt) = I + Qt + Q^2 \frac{t^2}{2!} + Q^3 \frac{t^3}{3!} + \cdots .\ee 
The standard technique for computing this exponential (at least in phylogenetics) is to first diagonalise the matrix $Q$ as 
\be Q = V \Lambda V^{-1}\ee 
where $\Lambda$ is a diagonal matrix, and then use the identity
\be \exp(Qt) = V \exp(\Lambda t) V^{-1},\ee 
noting that $\exp(\Lambda t)$ is a diagonal matrix with values $\exp(\Lambda_{ii} t)$ down the diagonal. 

\begin{example}
In the binary symmetric model, $Q = \begin{rmatrix} -1 & 1 \\ 1 & -1 \end{rmatrix}$. We have
\begin{eqnarray*}
Q &=& \begin{rmatrix} 1 & 1 \\ 1 & -1 \end{rmatrix} \begin{rmatrix} 0 & 0 \\ 0 & -2 \end{rmatrix} \begin{rmatrix} 1 & 1 \\ 1 & -1 \end{rmatrix}^{-1} 
\end{eqnarray*}
giving
\begin{equation} \label{eqn:binaryP}
P(t) =  \begin{rmatrix} \shalf + \shalf e^{-2 t} & \shalf - \shalf e^{-2 t} \\ \shalf - \shalf e^{-2 t} & \shalf + \shalf e^{-2 t} \end{rmatrix}.
\end{equation} 
These probabilities are often presented with $t$ scaled by a constant $\mu$. Here we assume that time has been scaled so that $\mu=1$. 
\end{example}

 \subsection{The $n$-taxa process}
 
We introduce an alternative way of describing the evolution of sequences along a tree. The main advantage of the approach is that the dependence on tree structure, and all of the complications that it introduces, can be encoded away.
 
 We will be working with a continuous time Markov chain on a much larger state space. Suppose that we have $n$ taxa. Consider the set of all vectors assigning a state (e.g. a binary value or a nucleotide) to every taxa. In the binary case there are $2^n$ of these vectors; in the $4$-state case there are $4^n$. The set of these vectors will be the state space of our process, however we will refer to these vectors as {\em values} of the process in order to minimise confusion with binary or nucleotide states.
 
Consider again the tree based model. At a particular time $t$, let $\bv_t$ denote the vector of states of the {\em ancestors} for each of the $n$ taxa. The {\em $n$-taxa process} is the continuous time Markov chain describing the evolution of these vectors.

\begin{example}
In figure~\ref{fig:modelseq} the initial value of the process is that the ancestors of all taxa have state $0$. Hence $\bv_{t_0} = [0,0,0,0]$. Between time $t_0$ and $t_1$ there is a substitution (on branch a). This occurs in the population that is ancestral to all of the taxa, so the effect is to change the value of the $n$-taxa process to $\bv_{t_1} = [1,1,1,1]$. On branch $b$ there is another substitution, though this is ancestral only to taxa $1$ and $2$. Thus, at time $t_2$ we have $\bv_{t_2} = [0,0,1,1]$. The process continues until, at the present time ($t_4$) the value of the process equals the observed states.
\end{example}

In a (slight) abuse of terminology we will say that a taxon is a {\em descendant} of a branch if it is a descendant of the population/ancestors represented by that branch. For example in  Figure~\ref{fig:modelseq} taxa 3 and 4 are the descendants of branch $c$.

The $n$-taxon process is a continuous time Markov chain, but it is not time-homogeneous. The transition probabilities depend on which lineages at a particular time are ancestral to which taxa. In the example, the rates of substitution for the $n$-taxa process will change at the time points $t_1,t_2,\ldots$. The process is homogeneous during the intervals between these time points. In what follows, we derive the rate matrices for the $n$-taxon process over each time interval.

We make the following notational conventions to lessen confusion between the different rate matrices involved.
\begin{enumerate}
\item We use $Q$ to denote the rate matrix for the underlying (binary symmetric or K3ST) substitution process  (binary symmetric or K3ST).
\item We use $\bQ^{(i)}$ to denote the rate matrix for the $n$-taxon process during time interval $[t_{i-1},t_i]$.
\item We use $\bR^{(b)}$ to denote the rate matrix for the $n$-taxon process restricted to substitutions occurring on a given branch $b$.
\end{enumerate}

We make extensive use of the {\em Kronecker product} of matrices. Given an $m \times n$ matrix $X$ and a $p \times q$ matrix $Y$ the Kronecker product (or {\em tensor product}) of $X$ and $Y$ is the $mp \times nq$ matrix
\be X \otimes Y = \begin{rmatrix} 
X_{11} Y & X_{12} Y & \cdots & X_{1n} Y \\
X_{21} Y & X_{22} Y & \cdots & X_{2n} Y \\
\vdots & & \ddots & \vdots \\
X_{m1} Y & \cdots & \cdots & X_{mn} Y 
\end{rmatrix}
\ee 
The elements of a Kronecker product can be indexed by vectors so, for example,
\be (X \otimes Y)_{[i,p],[j,q]} = X_{ij}Y_{pq}, \hspace{1cm} 
(X \otimes Y \otimes Z)_{[i,p,s],[j,q,t]} = X_{ij} Y_{pq} Z_{st}.\ee 
See \cite{Horn94} for a detailed introduction to the Kronecker Transform. 
We will make use of the following properties
\begin{lemma} Let $W,X,Y,Z$ be matrices with appropriate dimensions.
\begin{enumerate}
\item $(X \otimes Y) \otimes Z = X \otimes (Y \otimes Z)$.
\item $(X \otimes Y) (W \otimes Z) = XW \otimes YZ$.
\item Suppose that $X,Y$ are non-zero. Then $X\otimes Y$ is diagonal if and only if $X$ and $Y$ are diagonal.
\end{enumerate}
\end{lemma}
We print matrices formed from Kronecker products in boldface.

\section{Transition probabilities for the $n$-taxon process: binary symmetric case}

\subsection{Substitutions down a single branch: rate matrix}

For the moment, consider only substitutions that occur along one particular branch $b$ in the tree. We ignore substitutions occurring at the same time along other branches. Substitutions occur along branch $b$ at rate $1$. These substitutions affect only the taxa that are descendants of the branch $b$; let $A$ be that set of taxa. A substitution along the branch corresponds to flipping the entries $\bv_t[i]$ for which $i \in A$. These are the only substitutions in the $n$-taxon process restricted to the branch, and these substitutions occur at rate $1$. Thus the rate matrix $\bR^{(b)}$ of this restricted process is given by
\be  \bR^{(b)}_{\bu \bv} = \begin{cases}
 1 & \mbox{ if $\bu[i] \neq \bv[i]$ exactly when $i \in A$;} \\
-1 & \mbox{ if $\bu = \bv $;} \\
0 & \mbox{ otherwise.}
\end{cases}
\ee 

\begin{example}
In the example in Figure~\ref{fig:modelseq} we have $Q = \begin{rmatrix} -1 & 1 \\ 1 & 1 \end{rmatrix}$.
All elements of $\bR^{(b)}$ are zero, except the diagonal elements (all $-1$) and the elements 
\be \bR^{(b)}_{[0,0,0,0],[1,1,0,0]}, \bR^{(b)}_{[0,0,0,1],[1,1,0,1]},\ldots,\bR^{(b)}_{[1,1,1,1],[0,0,1,1]},\ee 
\be \bR^{(b)}_{[1,1,0,0],[0,0,0,0],}, \bR^{(b)}_{,[1,1,0,1],[0,0,0,1]},\ldots,\bR^{(b)}_{[0,0,1,1],[1,1,1,1]},\ee 
which all equal $1$.
\end{example}

We re-express $\bR^{(b)}$ in terms of the Kronecker product of simple $2 \times 2$ matrices. Define $E = \begin{rmatrix} 0 & 1 \\ 1 & 0 \end{rmatrix}$. 

\begin{lemma} \label{lem:Rbform}
Let $A$ be the set of taxa that are descendants of the population represented by branch $b$. For each $i=1,2,\ldots,n$ set $M^{(i)} = E$ if $i \in A$ and $M^{(i)} =I$ otherwise. Then
\be \bR^{(b)} = M^{(1)} \otimes M^{(2)} \otimes \cdots \otimes M^{(n)} - \bI.\ee 
\end{lemma}

\begin{example}
Consider branch $c$ in figure~\ref{fig:modelseq}. As $A=\{3,4\}$ we have 
\be \bR^{(c)} = I \otimes I \otimes E \otimes E - \bI.\ee 
\end{example}

\subsection{Substitutions down a single branch: transition probabilities} \label{sec:singlebranch}

In this section we show how to diagonalise the matrices $\bR^{(b)}$. One of the attractions of the Kronecker product is that we can generally obtain a diagonalisation of the product matrix in terms of its factors.

Let $H := H^{(1)}= \begin{rmatrix} 1 & 1 \\ 1 & -1 \end{rmatrix}$
and $\Lambda = H E H^{-1} = \begin{rmatrix} 1 & 0 \\ 0 & -1 \end{rmatrix}$. Thus $HEH^{-1}$ and $HIH^{-1}$ are both diagonal. We use Kronecker product to construct a matrix that diagonalises $\bR^{(b)}$.

The $n^{\th}$ order {\em Hadamard} matrix is defined 
\be \bH^{(n)} = H \otimes H \otimes \cdots \otimes H,\ee 
an $n$-fold Kronecker product. Note that $(\bH^{(n)})^{-1} = 2^{-n} \bH^{(n)}$.

\begin{lemma} \label{lem:RbDiagonal}
Let $\bH = \bH^{(n)}$, the $n^{\th}$ order Hadamard matrix, and let $\bR^b$ be the rate matrix for the $n$-taxon process restricted to branch $b$, binary symmetric case. Let $A$ be the set of taxa that are descendants of branch $b$. Then
\be \bLambda^{(b)}:=\bH \bR^{(b)} \bH^{-1}\ee 
is a diagonal matrix, with
\be \bLambda^{(b)}_{\bu \bu} = (-1)^{|\{i \in A : \bu[i] = 1\}|} -1\ee 
for all state vectors $\bu$.
\end{lemma}
\begin{proof}
Define matrices $M^{(i)}$ as in Lemma~\ref{lem:Rbform}. Then 
\be \bH (\bR^{(b)} + \bI) \bH^{-1} = (H M^{(1)} H^{-1}) \otimes \cdots \otimes (H M^{(n)} H^{-1}).\ee 
If $i \in A$ then $H M^{(i)} H^{-1} = H E H^{-1} = \Lambda$ while if $i \not \in A$ we have $H M^{(i)} H^{-1} = I$. The Kronecker product of diagonal matrices is diagonal, so $\bLambda$ is diagonal. 

For the diagonal values, note that 
\be \bLambda^{(b)}_{\bu \bu} = \prod_{i=1}^n (HM^{(i)}H^{-1})_{\bu[i] \bu[i]}\ee 
and that 
\be (HM^{(i)}H^{-1})_{\bu[i] \bu[i]} = \begin{cases} -1 & \mbox{ if $i \in A$ and $u[i]=1$ ;} \\ 1 & \mbox{ otherwise.} \end{cases}\ee 
\end{proof}

The transition probabilities down branch $b$ now follow directly from the diagonalisation, since
\be \exp(\bR^{(b)} t) = \bH^{-1} \exp(\bLambda^{(b)}) \bH\ee 
and $\exp(\bLambda^{(b)})$ is a diagonal matrix with entries $\exp(\bLambda^{(b)}_{\bu \bu})$ down the diagonal.

\subsection{Transition probabilities over multiple lineages}

During the intervals between time points there will be, in general, several lineages evolving independently. Because of this independence, the rate matrix $\bQ^{(i)}$ for the substitution process over all lineages is simply the sum of the rate matrices $\bR^{(b)}$ for the individual branches present at that time point. 

\begin{example}
In figure~\ref{fig:modelseq} the rate matrix for the $n$-taxon process between $t_2$ and $t_3$ equals 
\be \bQ^{(3)} = \bR^{(d)} + \bR^{(e)} + \bR^{(c)}\ee 
 as branches $d,e,c$ are present during this interval.
 \end{example}

Between time points $t_0$ and the present time $t_k$ we therefore have a sequence of rate matrices $\bQ^{(1)},\bQ^{(2)},\ldots,\bQ^{(k)}$. Each rate  matrix $\bQ^{(i)}$ equals the sum of the rate matrices $\bR^{(b)}$ for all branches $b$ present during the interval $[t_{i-1},t_i]$. During each interval, the probability transitions are given by the standard exponential formula
\be \bP^{(i)} =\exp\left({\bQ^{(i)} (t_i - t_{i-1})}\right)\ee 
so the transition probabilities between time $t_0$ and time $t_k$ are given by
\begin{eqnarray}
\bP &=& \bP^{(1)} \bP^{(2)} \cdots \bP^{(k)} \nonumber \\
&=& e^{\bQ^{(1)} (t_1 - t_{0})} e^{\bQ^{(2)} (t_2 - t_{1})} \cdots e^{\bQ^{(k)} (t_k - t_{k-1})} \label{eqn:expprod}
\end{eqnarray}

Now we make a critical simplification. The rate matrices $\bR^{(b)}$ down each branch are all diagonalised by the Hadamard matrix $\bH$. Hence so are the sums $\bQ^{(i)}$. Since every matrix $\bQ^{(i)}$ in the product \eqref{eqn:expprod} is diagonalised by the same matrix $\bH$, the rate matrices $\bQ^{(i)}$ all commute. If two matrices $\bX$ and $\bY$ commute then $\exp(\bX) \exp(\bY) = \exp(\bX+\bY)$. Applying this identity to \eqref{eqn:expprod} gives
\be \bP = \exp\left( \sum_{i=1}^k \bQ^{(i)}(t_i - t_{i-1})\right).\ee 

Now examine the sum $\sum_{i=1}^k \bQ^{(i)}(t_i - t_{i-1})$, a linear combination of the individual branch rate matrices $\bR^{(b)}$. The coefficient of each matrix $\bR^{(b)}$ is equal to the total length of time that the branch is present: the length of the branch. Let  $\bt$ denote the vector of branch lengths. We have now established the following theorem.

\begin{theorem} \label{thm:binary}
Let $\bP[\bt]$ be the matrix of transition probabilities in the $n$-taxon process for the binary symmetric case given a branch length vector $\bt$.
Define 
\be \bQ[\bt] = \sum_b \bR^{(b)} \bt_b \ee 
where $b$ ranges over branches in the tree, $\bR^{(b)}$ is the matrix given in Lemma~\ref{lem:Rbform}, and $\bt_b$ is the length of branch $b$. Then $\bH \bQ[\bt] \bH^{-1}$ is a diagonal matrix and
\begin{equation} \bP[\bt] = \exp({\bQ[\bt]})
\label{eqn:eQt}
\end{equation}
 \end{theorem}

The probability distribution for a tree can be recovered from \eqref{eqn:eQt} by noting that, at the root, the process is $\mathbf{0} = [0,0,\ldots,0]'$ with probability $\pi_0 = 1/2$ and $\mathbf{1} = [1,1,\ldots,1]'$ with probability $1/2$. If $\bu$ is the pattern of states at the leaves, then the probability of observing $\bu$ equals
\be  \bp = \frac{1}{2} \bP_{\mathbf{0} \bu}[\bt] + \frac{1}{2} \bP_{\mathbf{1}\bu}[\bt].\ee 

Interestingly,  \eqref{eqn:eQt} also applies to the case when there is not a single common ancestor for the taxa, a feature that may well prove useful in population genetics applications.

\subsection{Recovering the Hadamard formula}

The Hadamard conjugation formula \cite{Hendy89,Swofford96} assumes that one taxon has all zero states and gives the probabilities for patterns on the remaining taxa. We can retrieve the formula almost directly from Theorem~\ref{thm:binary}, giving a new proof for the Hadamard conjugation. This new derivation explains why the zero entry of the vector $\bq$ in \cite{Hendy89} is chosen to make the sum of all entries zero: the vector $\bq$ is simply a row out of the rate matrix $\bQ$.

\begin{theorem} \cite{Hendy89} 
Suppose that the tree has taxa at the root with state $0$. For each non-zero vector $\bu$ (indexed by the remaining taxa) let $\bq_\bu$ be the length of the branch with descendants $\{i:\bu[i] = 1\}$, if there is such a branch in the tree, and zero otherwise. Let $\bq_{\mathbf{0}}$ be the negative of the sum of all the branch lengths in the tree. Let $\bp_{\bu}$ be the probability of observing the pattern $\bu$ at the leaves. Then
\be \bp = \bH^{-1} \exp(\bH \bq).\ee 
Here the exponential is entry-wise.
\end{theorem}
\begin{proof}
Let $\mathbf{0}$ denote the vector $[0,0,\ldots,0]'$. We seek the probabilities $\bp_\bu = \bP_{\mathbf{0} \bu}[\bt]$. As $\bP[\tau]$ is symmetric, $\bP_{\mathbf {0} \bu} [\bt]= \bP_{\bu \mathbf{0}}[\bt]$. 

The vector $\bq$ is the  $\mathbf{0}$-column of $\bQ[\bt]$ Let $\bLambda = \bH \bQ[\bt] \bH^{-1}$, so that $\bH \bQ = \bLambda \bH$. The $\mathbf{0}$-column of $\bH$ is all ones, so the $\mathbf{0}$-column of $\bLambda \bH$ is made up of the diagonal entries of $\bLambda$. Hence the entries in $\bH \bq$ are the entries along the diagonal of $\bLambda$. Taking entry-wise exponentials, we have that $\exp(\bH \bq)$ equals the entries along the diagonal of $\exp(\bLambda)$ and so $\exp(\bH \bq)$ is the first column of $\exp(\bLambda) \bH$. The formula now follows from that fact that $\bP_{\bu \mathbf{0}}[\bt]$ is the $\mathbf{0}$-column of 
\begin{eqnarray*}
\bP[\bt] & = & \bH e^{\bLambda} \bH^{-1} \\
& = &  2^{-(n-1)} \bH e^{\bLambda} \bH \\
& = & \bH^{-1} e^{\bLambda} \bH.
\end{eqnarray*}
\end{proof}

\section{Transition probabilities for the $n$-taxon process: K3ST model}

We now extend the results of the previous sections to the K3ST model. In the interests of brevity we only outline the key steps in the derivation.

\subsection{Substitutions down a single branch: rate matrix}

Under the K3ST model there are three types of substitution. Instead of defining one matrix $E$ as above, we define three matrices
\be E_{I} = \begin{rmatrix} 0 & 1 & 0 & 0 \\ 1 & 0 & 0 & 0 \\ 0 & 0 & 0 & 1 \\ 0 & 0 & 1 & 0 \end{rmatrix} \hspace*{1cm}
 E_{II} = \begin{rmatrix} 0 & 0 & 1 & 0 \\ 0 & 0 & 0 & 1 \\ 1 & 0 & 0 & 0 \\ 0 & 1 & 0 & 0 \end{rmatrix} \hspace*{1cm}
 E_{III} = \begin{rmatrix} 0 & 0 & 0 & 1 \\ 0 & 0 & 1 & 0 \\ 0 & 1 & 0 & 0 \\ 1 & 0 & 0 & 0 \end{rmatrix} 
,\ee 
so that $Q = \alpha E_I  + \beta E_{II} + \gamma E_{III} - (\alpha + \beta + \gamma) I$.

\begin{lemma} \label{lem:Rbform4}
Let $A$ be the set of taxa that are descendants of the population represented by branch $b$. For each $i=1,2,\ldots,n$ set $M_I^{(i)} = E_I$ if $i \in A$ and $M^{(i)} =I$ otherwise. Likewise for $M^{(i)}_{II}$ and $M^{(i)}_{III}$.  Then
\be \bR^{(b)} = \alpha M_I^{(1)} \otimes \cdots \otimes M_I^{(n)} + \beta M_{II}^{(1)} \otimes \cdots \otimes M_{II}^{(n)} + \gamma M_{III}^{(1)} \otimes \cdots \otimes M_{III}^{(n)} - \bI.\ee 
\end{lemma}

The matrix $\bR^{(b)}$ is indexed by vectors of states. We number the states $0,1,2,3$ corresponding to $A,C,G,T$ respectively.

\subsection{Substitutions down a single branch: transition probabilities}

We use the same trick as before to diagonalise the rate matrix $\bR^{(b)}$ in the K3ST case: using the properties of the Kronecker product. Let 
\be H=H^{(2)} = \begin{rmatrix} 1 & 1 & 1 & 1 \\ 1 & -1 & 1 & -1 \\ 1 & 1 & -1 & -1 \\ 1 & -1 & -1 & 1 \end{rmatrix}\ee 
then define $\Lambda_I = H^{-1} E_I H$, $\Lambda_{II} = H^{-1} E_{II} H$ and $\Lambda_{III} = H^{-1} E_{III} H$. Then $\Lambda_I = diag(1,-1,1,-1)$, $\Lambda_{II} = diag(1,1,-1,-1)$ and $\Lambda_{III} = diag(1,-1,-1,1)$.

For this case, we let $\bH$ denote the $n$-fold product $H \otimes H \otimes \cdots \otimes H$, which is equal to the $2n^{\th}$ order Hadamard matrix.

\begin{lemma} \label{lem:RbDiagonalK3ST}
Let $\bH = \bH^{(2n)}$, the $2n^{\th}$ order Hadamard matrix, and let $\bR^b$ be the rate matrix for the $n$-taxon process restricted to branch $b$, K3ST case. Let $A$ be the set of taxa that are descendants of branch $b$. Then
\be \bLambda^{(b)}:=\bH \bR^{(b)} \bH^{-1}\ee 
is a diagonal matrix, with
\be \bLambda^{(b)}_{\bu \bu} = \alpha (-1)^{|\{i \in A : \bu[i] = \mbox{$1$ or $3$} \}|} + \beta (-1)^{|\{i \in A : \bu[i] = \mbox{$2$ or $3$} \}|} 
+ \gamma  (-1)^{|\{i \in A : \bu[i] = \mbox{$1$ or $2$} \}|}
-1\ee 
for all state vectors $\bu$.
\end{lemma}

The transition probabilities down branch $b$ now follow directly from the diagonalisation, since
\be \exp(\bR^{(b)} t) = \bH^{-1} \exp(\bLambda^{(b)}) \bH\ee 
and $\exp(\bLambda^{(b)})$ is a diagonal matrix with entries $\exp(\bLambda^{(b)}_{\bu \bu})$ down the diagonal.

\subsection{Transition probabilities over multiple lineages}

The progression from rate matrices for branches to rate matrices for the entire $n$-taxon process is almost identical in the K3ST case as in the binary symmetric model case. During each time interval $[t_{i-1},t_i]$ the rate matrix $\bQ^{(i)}$ for the $n$-taxon process is the sum of the rate matrices $\bR^{(b)}$ for branches present at that time. They are all diagonalised by the $2n^{\th}$ order Hadamard matrix $\bH$, so commute, and we have
\be \bP  =  \exp\left(\sum_{i=1}^k \bQ^{(i)} (t_i - t_{i-1})\right).\ee 
Furthermore, given the vector $\bt$ of branch lengths we have
\be \sum_{i=1}^k \bQ^{(i)} (t_i - t_{i-1}) = \sum_b \bR^{(b)} \tau_b,\ee 
establishing the following analogue to Theorem~\ref{thm:binary}.

\begin{theorem} \label{thm:K3ST}
Let $\bP[\bt]$ be the matrix of transition probabilities in the $n$-taxon process for the binary symmetric case given a branch length vector $\bt$.
Define 
\be \bQ[\bt] = \sum_b \bR^{(b)} \bt_b \ee 
where $b$ ranges over branches in the tree, $\bR^{(b)}$ is the matrix given in Lemma~\ref{lem:Rbform4}, and $\bt_b$ is the length of branch $b$. Then $\bH \bQ[\bt] \bH^{-1}$ is a diagonal matrix and
\be  \bP[\bt] = \exp({\bQ[\bt]})\ee 
 \end{theorem}

\bibliographystyle{plain}

\end{document}